\documentclass[amsmath,amssymb,amsbsy,reprint,prb,preprintnumbers,showpacs,superscriptaddress]{revtex4-2}
\usepackage{graphicx,color}
\usepackage{dcolumn}
\usepackage{bm}
\usepackage{braket}
\usepackage{mathtools}
\usepackage{ulem}
\usepackage[breaklinks,colorlinks=true,linkcolor=blue,urlcolor=blue,citecolor=blue]{hyperref}
\usepackage{times}
\usepackage{physics}
\usepackage{latexsym}
\usepackage{amsmath, amssymb}
\usepackage{mathtools}
\usepackage{multirow}

\newcommand{\be}{\begin{eqnarray}}
\newcommand{\ee}{\end{eqnarray}}

\begin{document}

\title{Hierarchy of emergent cluster states by measurement from symmetry-protected-topological states with large symmetry to subsystem cat state}
\date{\today}
\author{Yoshihito Kuno} 
\thanks{These authors equally contributed}
\affiliation{Graduate School of Engineering Science, Akita University, Akita 010-8502, Japan}
\author{Takahiro Orito}
\thanks{These authors equally contributed}
\affiliation{Institute for Solid State Physics, The University of Tokyo, Kashiwa, Chiba, 277-8581, Japan}
\author{Ikuo Ichinose} 
\affiliation{Department of Applied Physics, Nagoya Institute of Technology, Nagoya, 466-8555, Japan}


\begin{abstract}
We propose {\it measurement-producing hierarchy} emerging among correlated states by sequential subsystem projective measurements. 
We start from symmetry-protected-topological (SPT) cluster states with a large symmetry and apply sequential subsystem projective measurements to them and find that generalized cluster SPT states with a reduced symmetry appear in the subsystem of the unmeasured sites.
That prescription finally produces Greenberger-Home-Zeilinger states with long-range order in the subsystem composed of periodic unmeasured sites of the original lattice. 
The symmetry-reduction hierarchical structure from a general large symmetric SPT cluster state is clearly captured by the measurement update flow in the efficient algorithm of stabilizer formalism. 
This approach is useful not only for the analytical search for the measured state but also for numerical simulation with a large system size. 
We also numerically verify the symmetry-reduction hierarchy by sequential subsystem projective measurements applied to large systems and large symmetric cluster SPT states.
\end{abstract}


\maketitle
\section{Introduction} 
Quantum measurement applied to a quantum many-body state leads to a change of the state, and sometimes the operation induces an exotic non-locally-correlated state due to the backaction of quantum measurement. 
In this sense, quantum measurement can be regarded as an important tool to operate quantum many-body systems. 

Recently, study in the interdisciplinary area of quantum information and condensed matter physics is progressing rapidly \cite{Wen_text}. 
As a recent hot topic, the interplay of measurements and quantum circuits on many-body systems induces many interesting many-body dynamics and leads to interesting non-trivial steady states depending on the setup of the circuits acting on many-body systems. 
In particular, random unitary circuits attract lots of attention.
The systems exhibit measurement-induced entanglement phase transitions, which have been extensively studied these days \cite{Li2018,Skinner2019,Li2019,Vasseur2019,Chan2019,Szyniszewski2019,Choi2020,Bao2020,Gullans2020,Jian2020,Zabalo2020,Sang2021,Sang2021_v2,Nahum2021,Sharma2022,Fisher2022_rev,Block2022,Liu,Richter2023,Sierant2023,Suzuki2023,Kumer2023}. 
As another topic, measurement-only circuits with a suitable choice of measurement operators and suitable application probabilities generate unconventional phases of matter. 
Through projective measurements without unitary evolution, various interesting many-body steady states emerge such as symmetry-protected topological (SPT) phases \cite{Lavasani2021,Klocke2022,KI2023}, topological orders \cite{Lavasani2021_2,Negari2023,KOI2024}, and non-trivial thermal and critical phases \cite{Ippoliti2021,Sriram2023,KOI2023,Lavasani2023,Zhu2023}. 

Furthermore, by preparing some entangled state called resource state, the application of measurements with suitable spatial patterns to that state produces a specific entangled state in the subsystems of unmeasured sites. 
This process can be applied to a quantum computation, which is called measurement-based quantum computation (MBQC) \cite{Raussendorf2001,Raussendorf2006,Briegel2009,Wei_rev2018}. 
Such a measurement approach to many-body states is applied not only to carry out a quantum computation but also to efficiently produce interesting states of matter in condensed matter physics.
Recently, in that direction of study, ``cat state'' with long-range order (LRO), SPTs, topological ordered states, fractons, and non-Abelian topological ordered states are efficiently prepared by some measurement procedure applying to some proper entangled states \cite{Raussendorf2005,Verresen2021_cat,Tantivasadakarn2022,Lee2022,Lu2023}. Also a proposal of a stochastic quantum circuit model appearing such a LRO state exists \cite{Angelidi_2023}. 
More recently, a transition to such a ``cat state" through measurements has been observed in a real experimental quantum device \cite{Chen2023_v1}. 

From the viewpoint of the current tendency of research explained above, we shall study the measurement-induced state generation in many-body states by using suitable projective measurements. 
We show an argument on that issue: From an initial generalized cluster SPT state with large symmetry, a sequential measurement to subsystems induces a series of generalized cluster SPT states with a reduced set of symmetries. That is, we see {\it measurement-reduction hierarchy}. 
After subsystem measurements of suitable times, the initial generalized cluster SPT state reaches a cat state 
(Greenberger-Home-Zeilinger (GHZ) state) on a subsystem as the final state. 
This flow of many-body states can be regarded as a generalization of the methods to produce a cat state with LRO, which was recently proposed in \cite{Verresen2021_cat,Tantivasadakarn2022,Lee2022,Lu2023}.

Furthermore, we find efficient feedback-unitary operation for arbitrary projective measurements. 
Due to the introduction of that feedback unitary, we obtain genuine generalized cluster SPT states and final GHZ state for any patterns of measurement outcomes. 
This approach can be regarded as an extended method proposed in \cite{Lu2023}.

As a result, we show a rich hierarchical structure from large symmetric SPTs by measurements. 
We expect that this approach can apply to various many-body quantum systems and induce various correlated quantum many-body states. 
This work gives concrete examples of the above phenomenon including the numerical verification for the analytical observations.

The rest of this paper is organized as follows. 
In Sec.~II, we explain generalized cluster spin models and their SPT ground states. 
These states are target states in this work.   
In Sec.~III, before proposing our main argument, we explain the state preparation scheme for general cluster SPT states by employing a recently-proposed prescription on a quantum circuit.  
In Sec.~IV, we show our main argument in this work. 
There, we first explain the most general argument on states emerging as a result of sequential genuine projective measurements in subsystems. 
Then, we show a few concrete examples by the analytical calculation using the measurement update in the efficient algorithm of the stabilizer simulation. 
In Sec.~V, we explain feedback unitary, which plays an important role for `erasing' glassy properties of  emergent SPTs. 
That is, the feedback unitary helps us to create clean hierarchical cluster SPTs as well as final GHZ states. 
In Sec.~VI, we show results of the numerical simulation by using the efficient algorithm of the stabilizer formalism \cite{Gottesman1997,Aaronson2004}, and corroborate our argument for large system sizes and large-symmetric cluster initial states.
Section VII is devoted to conclusion.


\section{Generalized cluster model}
\begin{figure*}[t]
\begin{center} 
\vspace{0.5cm}
\includegraphics[width=16cm]{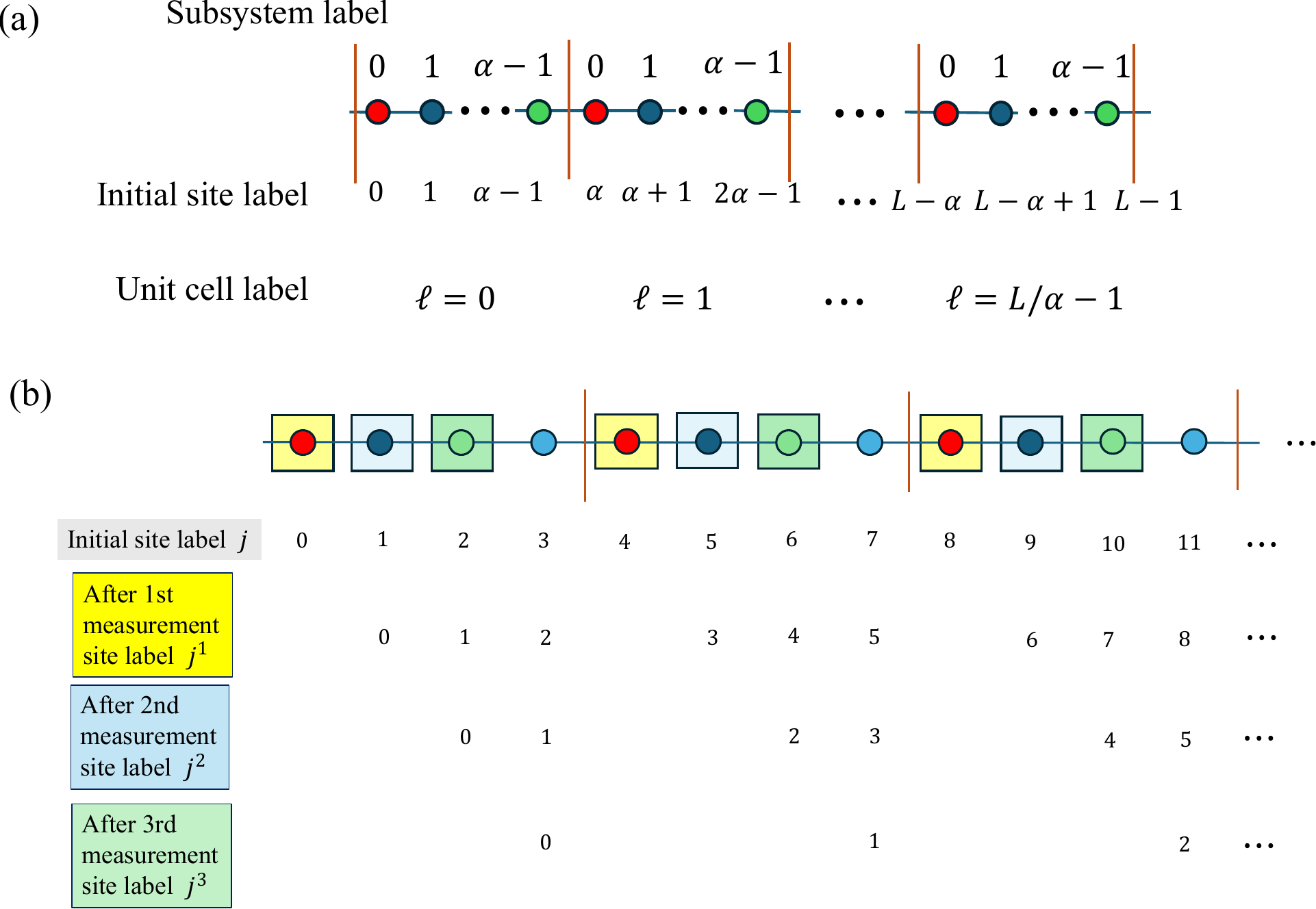}  
\end{center} 
\caption{
(a) Set up of the one-dimensional many-qubit system with periodic boundary conditions. 
One unit cell includes $\alpha$-different subsystem sites, $0,1,\cdots, (\alpha-1)$. 
The red sites are included in the subsystem $(sys)_{0}$, the dark blue ones in the subsystem $(sys)_{1}$, and the green ones in the subsystem $(sys)_{\alpha-1}$. 
The label $j$ denotes the original site label. 
The label $j$ is represented by $j=\alpha \ell + k$, where $\ell$ is the unit cell label and $k$ is the subsystem label. 
(b) Rule for labeling site after one measurement step.
Here, we show $\alpha=4$ example. 
Three measurement steps are considered. 
The unmeasured sites are relabeled in novel consequent order at each measurement step, the representation is denoted by $j^{m_s}$ with $m_s$. 
The renumbering of unmeasured sites after a measurement is used for the site-definitions of effective Hamiltonian and stabilizer generators, and the order parameters such as STO and SG. 
The rule of site labeling is shown in Appendix.C.}
\label{Figlabel}
\end{figure*}

This work focuses on the evolution of the ground state by local measurements, Hamiltonian of which is given by the following generalized cluster spin model \cite{Suzuki1971,Bartlet2009,Smacchia2011,Giampaolo2015,Lahtinen2015}, 
\begin{eqnarray}
H_{\rm gc}(\alpha)=-\sum^{L-1}_{j=0}Z_{j}\biggr[\prod^{\alpha-1}_{\ell=1}X_{j+\ell}\biggl]Z_{j+\alpha},
\label{gc_model}
\end{eqnarray}
where $Z_j$, $X_j$ are Pauli operators and $\alpha$ is an integer larger than 2. 
Hereafter, we call the above site-label $j$ initial site label, as shown in Fig.~\ref{Figlabel} (a).
We mostly employ periodic boundary conditions, that is, the system is a ring composed of $L$ qubits.

The above $\alpha$-cluster model has $\alpha$-global symmetries generated by the following operators \cite{Morral-Yepes2023}: 
\begin{eqnarray}
G^{X,\alpha}_{m}=\prod^{L/\alpha-1}_{\ell=0}X_{\alpha \ell+ m},
\label{symmetry}
\end{eqnarray}
where $m=0, 1, \cdots, \alpha-1$.
For any even $\alpha$, the ground state is the unique gapped SPT state protected by the $\alpha$-global $Z_2$ symmetries, $(Z_2)^{\alpha}$, corresponding to $ G^{X,\alpha}_{m}$ ($m=0,1,\cdots, \alpha-1$). 
The most familiar example is $\alpha=2$ case, the ground state of $H_{\rm gc}(2)$ is the cluster state protected by $Z_2\times Z_2$ symmetry\cite{Son2011}. 

On the other hand for any odd $\alpha$, 
the ground state of $H_{\rm gc}(\alpha)$ is doubly degenerate as a result of spontaneous symmetry breaking (SSB) \cite{Verresen2017},where the broken symmetry is the diagonal symmetry generated by $\prod_{m=0}^{\alpha-1}G_m^{X,\alpha}\equiv P$ corresponding to the parity operator.
Then, each state is a cluster SPT state protected by the $(Z_2)^{\alpha-1}$ global symmetries that remain under the SSB of the diagonal symmetry ~\cite{Verresen2017}. 
The two-fold degenerate odd-$\alpha$ cluster SPT states are to be distinguished by the sign of the parity operator 
$P\equiv \prod^{L-1}_{j=0} X_j$
[An example is given in \cite{Klocke2022}].  
For example for the $\alpha=3$ $(ZXXZ)$ model, the ground states are two distinct orthogonal states, each of which corresponds to a cluster SPT state protected by a global $Z_2\times Z_2$ symmetry ~\cite{Verresen2017} generated by  $G^{X,3}_{0}G^{X,3}_{1}$ and $G^{X,3}_{1}G^{X,3}_{2}$. 
The SSB for each state can be characterized by a correlator of a local order parameter $Z_jY_{j+1}Z_{j+2}$ \cite{Morral-Yepes2023}.  Here, we comment that this system plays an important role in quantum computation and quantum error correcting codes \cite{Klocke2022}. 
For the specific $\alpha=1$ case, the model is nothing but the Ising model without a transverse field and the ground states are doubly-degenerate $L$-site GHZ states with distinct parity $P=\pm 1$.

\section{State preparation for generalized cluster SPT states}
Before going to the main findings of this work, we discuss methods of the state preparation for our target generalized $\alpha$ cluster SPT states. 
Readers who are interested only in the main results obtained in this work can skip to Sec.~IV.

Generalized $\alpha$ cluster SPT states under consideration can be prepared from a simpler state by using the combination of sequential controlled-Z gates (CZ gates), which is sometimes called cluster entangler \cite{Raussendorf2001,Tantivasadakarn2022} and defined by 
\begin{eqnarray}
U_{CZ}\equiv\prod^{L-1}_{j=0}CZ_{j,j+1},
\end{eqnarray}
where $CZ_{j,j+1}$ represents CZ gate for nearest neighbor sites, $j$ and $(j+1)$. 
More generally, various cluster SPT states are to be generated by the pivot transformation~\cite{Tantivasadakarn2023}.
The pivot transformation by $h^{j}_{k}$ is defined as
\begin{eqnarray}
U^{p}_{k}=\exp\biggr(i\frac{\pi}{4}\sum_j h^{k}_{j}\biggl),
\end{eqnarray}
where $h^{k}_j$ are given by 
$h^{k}_{j}=Z_jX_{j+1}\cdots X_{j+k-1}Z_{j+k}$ with $k>0$.
In general, the pivot transformation induces the following formula, which we shall use in the following analysis: 
\begin{eqnarray}
U^{p}_{k_0}h^{k}_jU^{p\dagger}_{k_0}=h^{2k_0-k}_{j+k-k_0}.
\end{eqnarray}

\noindent \underline{Even $\alpha$ case}:
We denote a general even-$\alpha$ cluster SPT state as $|CS_e(\alpha)\rangle$, which is the \textit{unique ground state state} of the Hamiltonian $H_{\rm gc}(\alpha)$. 
First, the $\alpha=2$ cluster SPT state can be created from the $+X$ product state (the unique state) denoted by $|+\rangle^{\otimes L}$ \cite{Raussendorf2001,Tantivasadakarn2022} as
\begin{eqnarray}
|CS_e(2)\rangle=U_{CZ}|+\rangle^{\otimes L}.
\label{state_pre1}
\end{eqnarray}

Based on the state $|CS_e(\alpha)\rangle$, the application of the pivot transformation $U^{p}_{r+2}$ to it creates a general $\alpha=2r+2$ ($r \in \mathbb{N}$) even-$\alpha$ cluster SPT state as 
\begin{equation}
|CS_{e}(\alpha)\rangle=U^{p}_{r+2}|CS_{e}(2)\rangle.
\label{state_pre12}
\end{equation}
This comes from the fact that the Hamiltonian $H_{\rm gc}(2)$ is transformed by the conjugation of the pivot transformation $U^{p}_{r+2}$ as $U^{p}_{r+2}H_{\rm gc}(2)U^{p\dagger}_{r+2} =H_{\rm gc}(2+2r)$.
(Please note that $H_{\rm gc}(2)$ and $U^p_2$ commute with each other.)
By this transformation, the ground state $|CS_e(2)\rangle$ is transformed into $|CS_e(2+2r)\rangle$. 
Please note that the uniqueness of the ground state is preserved in this transformation.

In this way, we can prepare any even-$\alpha$ cluster SPT state from the simple product state.\\

\noindent \underline{Odd-$\alpha$ case}:
Next, let us discuss the preparation of a general odd-$\alpha$ cluster SPT state. 
Note that the ground state is two-fold degenerate in this case \cite{Verresen2017,Morral-Yepes2023}. 
In this work, we mostly focus on one of the degenerate ground states, an eigenstate of the parity $P=\prod^{L-1}_{j=0}X_j$, which is a logical operator from the quantum information point of view.
We start from one of the GHZ ground states of the quantum Ising Hamiltonian $H_{ZZ}=-\sum_{j}Z_jZ_{j+1}$, i.e., the ground state with even parity $P=+1$ such as 
$|{\rm GHZ}_{+}\rangle=\frac{1}{\sqrt{2}}(|\uparrow\rangle^{\otimes L}+|\downarrow\rangle^{\otimes L})$. 
From the state $|{\rm GHZ}_{+}\rangle$, application of the pivot transformation $U^{p}_{r+1}$ creates a general $\alpha=2r+1$ ($r \in \mathbb{N}$) cluster SPT state as 
\begin{equation}
|CS_{o}(\alpha)\rangle=U^{p}_{r+1}|{\rm GHZ}_{+}\rangle.
\label{state_pre11}
\end{equation}
It is straightforward to show that the resultant state $|CS_{o}(\alpha)\rangle$ has positive parity $P=+1$.

Here, we remark that the pivot transformation for arbitrary $k$ can be implemented by a combination of quantum gates on the quantum circuit. 
Therefore, by using the cluster entangler and the pivot transformation, we can prepare any even and odd-$\alpha$ cluster SPT states from the two kinds of states $|+\rangle^{\otimes L}$ and $|{\rm GHZ}_{+}\rangle$, respectively.

\section{Sequential subsystem measurements for a general cluster SPT state}

In this section, we shall give a qualitative discussion on states emerging through sequential subsystem measurements starting from cluster SPT states, the ground states of $H_{\rm gc}(\alpha)$ for various $\alpha$'s.
Then, we show two concrete examples in small systems by using the analytically tractable update methods in the efficient algorithm of stabilizer formalism. 
Furthermore, by making use of suitable feedback unitary incorporating information of outcomes \cite{Lu2023}, we show that the `hierarchical structure' of the resultant states appears for any measurement outcomes. 
(The details will be discussed in the subsequent section.)
There, entanglement, topological properties, and symmetries exhibit interesting behavior under sequential local measurements.
This is one of the main findings of this study.

\subsection{General argument}
\noindent\underline{General argument for even-$\alpha$ case}: 
We first consider a general even-$\alpha$ case.
The $\alpha$ cluster SPT pure state $|CS_e(\alpha)\rangle$ is defined on the ring with length $L=\alpha N$, where $L$ is the total number of sites with periodic boundary conditions and $N$ is the number of unit cells. 
As shown in Fig.~\ref{Figlabel} (a), the site label $j$ is the initial site label, $j=0,1,\cdots L-1$, and we introduce $\alpha$ subsystems, which have $L/\alpha$ lattice sites. 
Here, sites in each $\alpha$ subsystem are numbered as $(sys)_{k}=\{ j=\alpha \ell +k | \ell=0,\cdots,L/\alpha -1\}$ for $k=0,1,\cdots, \alpha-1$, where $\ell$ numbers unit cells and $k$ labels internal sites (corresponding to the subsystem label) in a unit cell. 
These schematics are also shown in Fig.\ref{Figlabel} (a).  

We consider to perform a sequence of one-layer projective measurements acting on all sites in the subsystem $(sys)_{k}$. 
The one-layer projective measurement operator is given by
\begin{eqnarray}
P^{k}_{{\vec \beta}^k}=\prod_{j\in (sys)_{k}}\frac{1+\beta_jX_j}{2}
\label{pro1}
\end{eqnarray}
where ${\vec{\beta}}^k=\{\beta_{0+k},\beta_{\alpha+k},\cdots, \beta_{\alpha(L/\alpha-1)+k}\}$
is a set of measurement outcomes defined on the subsystem $(sys)_{k}$ corresponding to the eigenvalue of $X_j$ with $\beta_j=\pm 1$. 

We firstly apply the measurement $P^{0}_{{\vec \beta}^0}$, that is, measure all of the sites of the subsystem $(sys)_{0}$. 
Here, we regard it as the first measurement step represented by $m_s=1$, where we introduce a label $m_s$ denoting the number of measurement steps.
Then, the initial state changes as follows,
\begin{eqnarray}
P^{0}_{{\vec \beta}^0}|CS_e(\alpha)\rangle \propto |CS^g_o(\alpha-1)\rangle \otimes |{\vec{\beta}}^0_x\rangle_{(sys)_{0}}.
\label{1st_md_state}
\end{eqnarray}
Here, $|CS^g_o(\alpha-1)\rangle$ is a glassy $(\alpha-1)$ cluster SPT state with a parity $P_{0}\equiv \prod_{j\in (all)-(sys)_{0}}X_j=1$ through non-trivial correlations between outcomes (please see the comments below Eq.~(\ref{al=4_m=1})),
where $(all)-(sys)_{0}$ denotes the set of all sites except the measured sites in $(sys)_{0}$ (the label $(all)$ denotes the set of all initial sites, the number of which is $L$). 
In addition, the fact $P_{0}\equiv \prod_{j\in (all)-(sys)_{0}}X_j=1$ gives an insight into finding a feedback unitary discussed in Sec.~V. 
The glassy state $|CS^g_o(\alpha-1)\rangle$ residing on the entire unmeasured sites is one of the two-fold degenerate ground state of the following effective Hamiltonian given as
\begin{eqnarray}
H^{\rm eff}(0)=
-\sum^{L-L/\alpha-1}_{j^1=0}\beta_{n^0(j^1)}Z_{j^1}\biggr[\prod^{\alpha-2}_{\ell=1}X_{j^1+\ell}\biggl]Z_{j^1+\alpha-1},\nonumber\\
\label{1st_md_state_effH}
\end{eqnarray}
where the unmeasured sites after the first step measurement are renumbered in order, and we denote them by $j^1$ as explicitly shown in Fig.\ref{Figlabel} (b) and for the labeling-rule between the initial site label $j$ and $j^1$ is given in Appendix C.
On the RHS of Eq.~(\ref{1st_md_state_effH}), the site label of outcome $n^{0}[j^1]$ 
denotes the measured site in the support of original operator $ZX\cdots XZ$ (stabilizer) to which the site $j^1$ belongs. The labeling-rule is given in Appendix C.
In terms of the stabilizer formalism \cite{Gottesman1997,Nielsen_Chuang}, the representation of the set of the stabilizer generator for the glassy state $|CS^g_o(\alpha-1)\rangle$ is given by $\mathcal{S}^{\alpha-1}=[g^{\alpha-1}_0, g^{\alpha-1}_1, \cdots, g^{\alpha-1}_{L-L/\alpha-1}, P_0]$, where $g^{\alpha-1}_{j^1}=\beta_{n^0(j^1)} Z_{j^1}\biggr[\prod^{\alpha-2}_{\ell=1}X_{j^1+\ell}\biggl]Z_{j^1+\alpha-1}$. 
On the other hand, the state $|\vec{\beta}_x\rangle_{(sys)_{0}}$ is a $X$-directed product state on the subsystem $(sys)_{0}$, where the directions depend on the set of outcomes ${\vec{\beta}}^0$. 
Herein, we see that the one-layer projective measurement $P^{0}_{\vec \beta}$ for the $\alpha$ cluster SPT state produces the $(\alpha-1)$(odd) cluster SPT state with $P_0=+1$ appearing on the unmeasured sites. 
We also comment that the outcome factors $\beta$'s in the effective Hamiltonian $H^{\rm eff}(0)$ can be eliminated by introducing a feedback unitary as it is discussed in Sec.V.

\begin{figure*}[t]
\begin{center} 
\includegraphics[width=18cm]{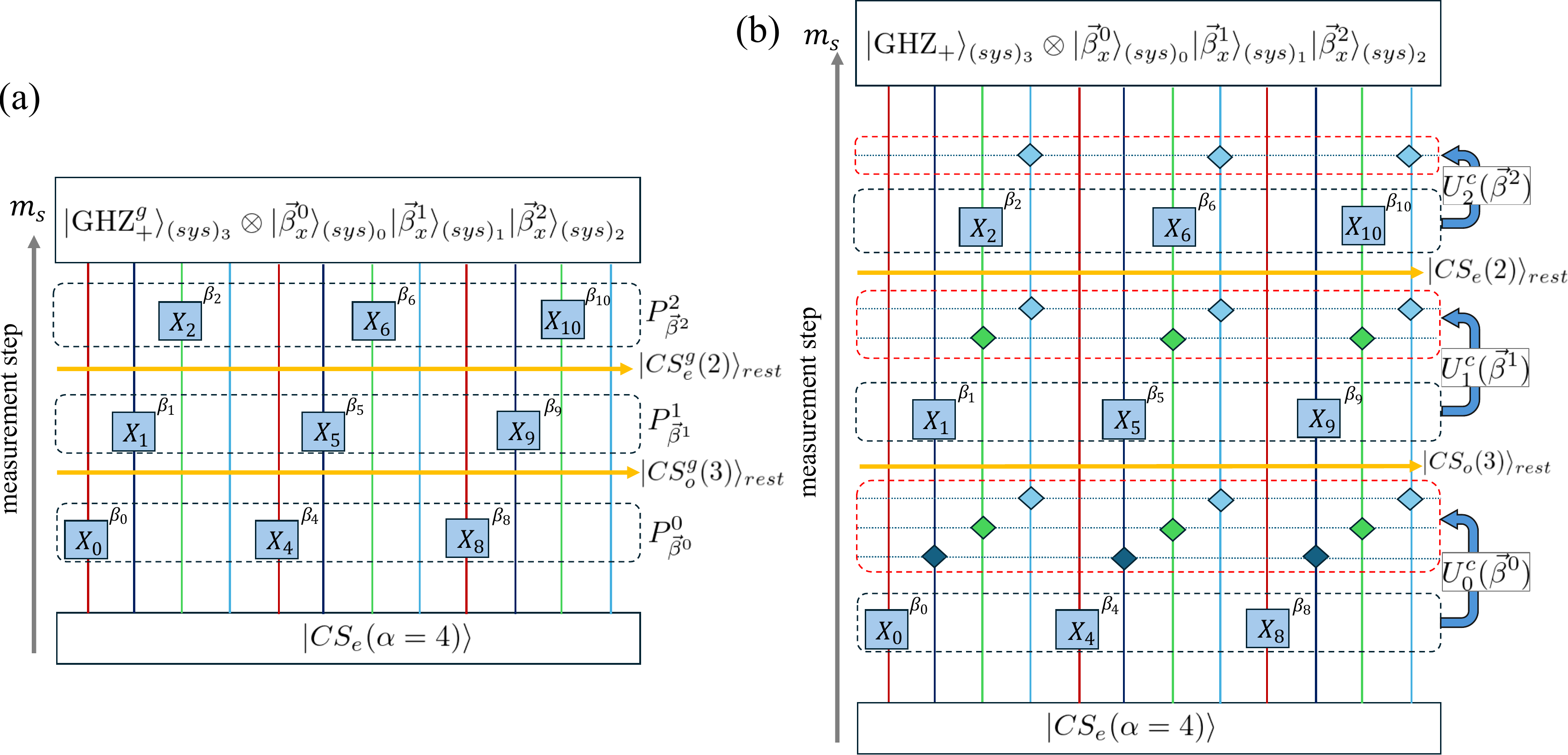}  
\end{center} 
\caption{
Schematic circuit picture for $\alpha=4$ and $N=3$ cases:
(a) Sequential measurements applied to initial $\alpha=4$ cluster SPT state. 
The three different measurement layers on different subsystems are applied. 
After each measurement step, the reduced cluster SPT states appears on the unmeasured sites and finally, after $(\alpha-1)=3$ measurement steps, the glassy GHZ state is produced on the unmeasured subsystem $(sys)_{3}$. 
(b) Measurement and feedback unitary prescription corresponding to LOCC. 
The red dotted blocks and diamond makers represent feedback unitaries. 
The line colors in the circuit represent each subsystem, where the red, dark blue, green, and right blue are $(sys)_{0}$, $(sys)_{1}$, $(sys)_{2}$ and $(sys)_{3}$, respectively. 
The right blue lines are the unmeasured sites at the final stage. 
Here, we observe the clean GHZ state with LRO on the subsystem $(sys)_{3}$.}
\label{Fig2}
\end{figure*}
We can easily expect that by employing the above manipulation in a sequential manner, we obtain a series of glassy cluster SPT states with reduced symmetries defined on the unmeasured sites. 
That is, as the second step ($m_s=2$), we further apply another one-layer projective measurement $P^{1}_{{\vec \beta}^1}$ to the above state to obtain outcomes $\vec{\beta}^1=(\beta^1_{1},\beta^1_{\alpha+1},\cdots, \beta^1_{\alpha(L/\alpha-1)+1})$ on $(sys)_{1}$, and then
\begin{eqnarray}
&&P^{1}_{{\vec \beta}^1}|CS^g_o(\alpha-1)\rangle \otimes |\vec{\beta}_x\rangle_{(sys)_{0}}\nonumber\\
&&\propto|CS^g_e(\alpha-2)\rangle \otimes |\vec{\beta}^0_x\rangle_{(sys)_{0}} |\vec{\beta}^1_x\rangle_{(sys)_{1}},
\label{2st_md_state}
\end{eqnarray}
where $|CS^g_e(\alpha-2)\rangle$ is a glassy $(\alpha-2)$ cluster SPT state. The state can be regarded as the unique ground state of the effective Hamiltonian defined on the remaining unmeasured all sites, given as
\begin{eqnarray}
&&H^{\rm eff}(1)=\nonumber\\
&&-\sum^{L-2L/\alpha-1}_{j^2=0}\beta_{n^0[j^1[(j^{2})^{-1}]]}\beta_{n^1(j^2)}Z_{j^2}\biggr[\prod^{\alpha-3}_{\ell=1}X_{j^2+\ell}\biggl]Z_{j^2+\alpha-2},\nonumber\\
\label{2st_md_state_effH}
\end{eqnarray}
where the unmeasured sites after the second step measurement are again labeled in order as $j^2=j^2[j]$ shown in Fig.~\ref{Figlabel} (b), (its inverse denotes $j=(j^2)^{-1}$) and the label of outcomes $n^{1}(j^2)$ is defined as in the previous case, see Appendix C, showing their labeling-rules. 
Here, we see that after the measurement $P^{1}_{\vec{\beta}^1}$, the $(\alpha-1)$-cluster SPT state is turned into the $(\alpha-2)$(odd) cluster SPT state appearing on the unmeasured sites. 
This indicates {\it reduction hierarchy}: the reduced symmetric glassy cluster SPT state appears on the unmeasured sites. 
This procedure results in inducing further small symmetric cluster SPT states on the unmeasured sites, and finally, after the $(\alpha-1)$-times one-layer measurements ($m_s=\alpha-1$) for each subsystem up to $(sys)_{\alpha-2}$, the final measured state comes to be   
\begin{eqnarray}
\biggl(\prod^{\alpha-2}_{k=0} P^{k}_{{\vec \beta}^k}\biggr)|CS_e(\alpha)\rangle \propto |{\rm GHZ}^g_+\rangle_{(sys)_{\alpha-1}} \bigotimes^{\alpha-2}_{k=0} |\vec{\beta}^{k}_x\rangle_{(sys)_{k}}.\nonumber\\
\label{last_md_state}
\end{eqnarray}
We see the above $\alpha$-period long-range-ordered state as the `glassy GHZ state' with $P_{\alpha-2}\equiv \prod_{j\in (all)-\sum^{\alpha-2}_{k=0}(sys)_{k}}X_j=1$ denoted by $|{\rm GHZ}^g_+\rangle_{(sys)_{\alpha-1}}$ defined on the subsystem $(sys)_{\alpha-1}$. 
We used the terminology  `glassy GHZ state' in the above as the orientation of spin at each unmeasured site varies depending on the outcomes but there still exists long-range entanglement in the resultant subsystem such as 
${1\over \sqrt{2}}(|\uparrow\uparrow\downarrow\cdots\rangle+|\downarrow \downarrow\uparrow\cdots\rangle)$. The concrete schematic picture of this approach is shown in Fig.~\ref{Fig2} (a).\\

\noindent\underline{General argument for odd $\alpha$ case}: 
Similarly for the general odd $\alpha$ case, we can apply the same procedure with the even $\alpha$ case.
We start from one of the $\alpha$-cluster SPT states with $P=+1$, defined on $L=\alpha N$, where $L$ is the total number of site with periodic boundary conditions and $N$ is the number of unit cells. 
Then, we first apply the one-layer projective measurement operator $P^{0}_{{\vec \beta}^0}$ to the initial state ($m_s=1$). 
The resultant state is  obtained as follows,
\begin{eqnarray}
P^{0}_{{\vec \beta}^0}|CS_o(\alpha)\rangle \propto |CS^g_e(\alpha-1)\rangle \otimes |\vec{\beta}^0_x\rangle_{(sys)_{0}},
\label{1st_md_state_odd}
\end{eqnarray}
where $|CS^g_e(\alpha-1)\rangle$ is a glassy $(\alpha-1)$(even) cluster SPT state corresponding to the unique ground state of the effective Hamiltonian defined on the remaining unmeasured all sites, 
\begin{eqnarray}
H^{\rm eff}_e(0)=
&&-\sum_{j^1}\beta_{n^0(j^1)}Z_{j^1}\biggr[\prod^{\alpha-2}_{\ell=1}X_{j^1+\ell}\biggl]Z_{j^1+\alpha-1},\nonumber\\
\label{1st_md_state_effH2}
\end{eqnarray}
where the unmeasured sites after the first step measurement are again labeled in order as $j^1$ previously explained in Fig.~\ref{Figlabel} (b) and Appendix C, and also the label of the outcome factor $n^{0}[j^1]$ is defined as already explained.

Then, we apply the second-step projective measurement operator $P^{1}_{{\vec \beta}^1}$ to the former one ($m_s=2$), 
\begin{eqnarray}
&&P^{1}_{\vec{\beta}^1}|CS^g_e(\alpha-1)\rangle \otimes |\vec{\beta}^0_x\rangle_{(sys)_{0}}\nonumber\\
&&\propto|CS^g_o(\alpha-2)\rangle \otimes |\vec{\beta}^0_x\rangle_{(sys)_{0}} |\vec{\beta}^1_x\rangle_{(sys)_{1}},
\label{2st_md_state}
\end{eqnarray}
where $|CS^g_o(\alpha-2)\rangle$ is a glassy $(\alpha-2)$(odd) cluster SPT state with the positive parity, $P_1=+1$.


We repeat this prescription. 
After the $(\alpha-1)$-times projective one-layer measurements ($m_s=\alpha-1$), the final state is the same as that of even $\alpha$ case, 
\begin{eqnarray}
\biggl(\prod^{\alpha-2}_{k=0} P^{k}_{{\vec \beta}^k}\biggr)|CS_o(\alpha)\rangle\propto|{\rm GHZ}^g_+\rangle_{(sys)_{\alpha-1}} \bigotimes^{\alpha-2}_{k=0} |\vec{\beta}^{k}_x\rangle_{(sys)_{k}}.\nonumber\\
\label{last_md_state2}
\end{eqnarray}

Consequently for any initial $\alpha$-cluster SPT state, suitable $(\alpha-1)$-layer projective measurements induce $\alpha$-period LRO exhibited by the glassy GHZ state.

\subsection{Concrete example I : $\alpha=4$ case in a small system}
In this subsection, we examine and verify the above prescription by analytical methods for small-size systems with small $\alpha$ using the handwritten update procedure of a set of stabilizer generators in the efficient numerical algorithm \cite{Gottesman1997,Aaronson2004}. 
Readers interested only in the verification of our argument for a large system size and large $\alpha$ case can skip to Sec.VI.
The basic transformation of the set of stabilizer generators and update procedure are explained in Appendices A and B. 
Observation of the stabilizer generators gives us lots of insight into the change of the system under projective measurements. 

First, as an example of even $\alpha$ system, we consider $L=\alpha N$ with $N=3$.
In the stabilizer formalism, the initial $\alpha=4$ cluster SPT state is characterized by a set of 12 stabilizer generators denoted by $\mathcal{S}^{\alpha=4}(m_s=0)$, which are given by 
$$
\mathcal{S}^{\alpha=4}(m_s=0)=[g^4_{0},\cdots, g^4_{11}],
$$
where 
$g^4_{j}$ is $j$-th stabilizer generator given by  $g^4_{j}=Z_jX_{j+1}X_{j+2}X_{j+3}Z_{j+4}$ (note that $j$ is the initial site label).

Let us apply $P^{0}_{{\vec \beta}^0}$, then we obtain a updated set of stabilizer generators, 
\begin{eqnarray}
&&\mathcal{S}^{\alpha=4}(m_s=0)\xlongrightarrow{P^{0}_{{\vec \beta}^0}}
\mathcal{S}^{\alpha=4}(m_s=1)\nonumber\\
&&=[\beta_0 X_{0},\beta_4 X_{4},\beta_8 X_{8},\nonumber\\
&&\beta_4g^{3}_1,\beta_4g^{3}_2,\beta_4g^{3}_3,
\beta_8g^{3}_5,\beta_8g^{3}_6,\beta_8g^{3}_7,\nonumber\\
&&\beta_0g^{3}_9,\beta_0g^{3}_{10}
,X_1X_2X_3X_5X_6X_7X_9X_{10}X_{11}],
\label{al=4_m=1}
\end{eqnarray}
where we have used the basic transformation among stabilizer generators several times and re-definitions them such as $g^{3}_{j}=Z_{j^1[j]}XXZ$ and also the last element of the parity $\prod X$, both of which are defined on the unmeasured sites.
Here, we should remark that the outcomes have the following strict correlation $\beta_8=\beta_0\beta_4$, which stems from the fact that the ground state of the $\alpha=4$ Hamiltonian has $P=+1$. 
From the above set of stabilizer generators, the resultant state is obtained as in the form of Eq.~(\ref{1st_md_state})
with the positive parity.

As the second step, the projective measurement $P^{1}_{{\vec \beta}^1}$ is applied as
\begin{eqnarray}
&&\mathcal{S}^{\alpha=4}(m_s=1)\xlongrightarrow{P^{1}_{{\vec \beta}^1}}
\mathcal{S}^{\alpha=4}(m_s=2)\nonumber\\
&&=[\beta_0 X_{0},\beta_4 X_{4},\beta_8 X_{8},
\beta_1 X_{1},\beta_5 X_{5},\beta_9 X_{9},\nonumber\\
&&\beta_4\beta_{5}g^{2}_2,\beta_4\beta_{5}g^{2}_3,
\beta_8\beta_{9}g^{2}_6,\beta_8\beta_{9}g^{2}_7,
\beta_0\beta_{1}g^{2}_{10},\beta_0\beta_{1}g^{2}_{11}
],\nonumber\\
\label{al=4_m=2}
\end{eqnarray}
where we have made use of the basic transformation for several times, $g^{2}_j=Z_{j^2[j]}XZ$, and we again find outcomes correlation such as $\beta_9=\beta_1\beta_5$. 
From this set of the stabilizer generators, we obtain the resultant state as
$|CS^g_e(2)\rangle \otimes |\vec{\beta}^0_x\rangle_{(sys)_{0}}\otimes |\vec{\beta}^1_x\rangle_{(sys)_{1}}$.

Finally, the last projective measurement $P^{2}_{\vec \beta}$ is applied as 
\begin{eqnarray}
&&\mathcal{S}^{\alpha=4}(m_s=2)\xlongrightarrow{P^{2}_{{\vec \beta}^2}}
\mathcal{S}^{\alpha=4}(m_s=3)\nonumber\\
&&=[\beta_0 X_{0},\beta_4 X_{4},\beta_8 X_{8},
\beta_1 X_{1},\beta_5 X_{5},\beta_9 X_{9},\nonumber\\
&&\beta_2 X_{2},\beta_6 X_{6},\beta_{10} X_{10},\nonumber\\
&&\beta_4\beta_{5}\beta_{6}g^{1}_3, 
\beta_0\beta_{1}\beta_{2}g^{2}_{11},X_3X_7X_{11}
],\nonumber\\
\label{al=4_m=3}
\end{eqnarray}
with the basic transformation, $g^{1}_j=Z_{j^3[j]}Z$, and the last element is the parity $P$ on the subsystem $(sys)_{3}$ obtained by the outcomes correlation $\beta_{10}=\beta_2\beta_6$. 
From this set of the stabilizer generators, the stabilizer state corresponds to $|{\rm GHZ}^g_+\rangle_{(sys)_{3}} \bigotimes^{2}_{k=0} |\vec{\beta}^{k}_x\rangle_{(sys)_{k}}$.

We conclude that we have verified the argument for the $\alpha=4$ case in the analytical level by using the update of the efficient algorithm for the stabilizer formalism.

\subsection{Concrete example II : $\alpha=3$ case in a small system} 

Here as an odd $\alpha$ example, we consider $L=N\alpha$ with $\alpha=N=3$. 
The same calculation as that of the former example can be applied. 
In fact, the present case is simpler than the former one. 
In the stabilizer formalism, the $\alpha=3$ cluster SPT state with $P=+1$ is given by an initial set of 9 stabilizer generators denoted by $\mathcal{S}^{\alpha=3}(m_s=0)$, given by 
$$
\mathcal{S}^{\alpha=3}(m_s=0)=[g^3_{0},\cdots, g^3_{7},X_0X_1X_2X_3X_4X_5X_6X_7X_8]
$$
where the last element requires that the state is in the $P=+1$ sector.

Let us apply $P^{0}_{{\vec \beta}^0}$, then we obtain the updated set of stabilizer generators, 
\begin{eqnarray}
&&\mathcal{S}^{\alpha=3}(m_s=0)\xlongrightarrow{P^{0}_{{\vec \beta}^0}}
\mathcal{S}^{\alpha=3}(m_s=1)\nonumber\\
&&=[\beta_0 X_{0},\beta_3 X_{3},\beta_6 X_{6},\beta_3g^{2}_1,\beta_3g^{3}_2,\beta_6g^{2}_4,
\beta_6g^{2}_5,\beta_0g^{2}_7,\beta_0g^{2}_8],\nonumber\\
\label{al=3_m=1}
\end{eqnarray}
where we have used the basic transformation several times and we find the outcomes have the correlation $\beta_6=\beta_0\beta_3$ coming from the positive parity $P=+1$ of the initial state. 
From this set of the stabilizer generators, the stabilizer state has the form of Eq.~(\ref{1st_md_state}). 

Further, the last projective measurement $P^{1}_{{\vec \beta}^1}$ is applied as 
\begin{eqnarray}
&&\mathcal{S}^{\alpha=3}(m_s=1)\xlongrightarrow{P^{1}_{{\vec \beta}^1}}
\mathcal{S}^{\alpha=3}(m_s=2)\nonumber\\
&&=[\beta_0 X_{0},\beta_3 X_{3},\beta_6 X_{6},
\beta_1 X_{1},\beta_4 X_{4},\beta_7 X_{7},\nonumber\\
&&\beta_3\beta_4g^{1}_2,
\beta_0\beta_1g^{1}_8,X_2X_5X_8],
\label{al=3_m=2}
\end{eqnarray}
with $g^{1}_j=Z_{j^2[j]}Z$, and the last element is the parity $P_2$ defined on the subsystem $(sys)_{2}$ (We also find the outcomes has the correlation $\beta_{7}=\beta_1\beta_4$.)
From this set of the stabilizer generators, the stabilizer state corresponds to $|{\rm GHZ}^g_+\rangle_{(sys)_{2}} \bigotimes^{1}_{k=0} |\vec{\beta}^{k}_x\rangle_{(sys)_{k}}$.

We have verified the argument for the $\alpha=3$ case in the analytical level. The case for larger system size and $\alpha$ is numerically verified in Sec.VI.

\section{Introducing feedback unitary}
So far, we have only considered applying the projective measurements $\{P^{k}_{{\vec \beta}^k}\}$ to the subsystem $(sys)_k$, and
focused on the output states depending on random measurement outcomes.
In other words, we have mostly observed measurement trajectories. 
However recently, a feedback operation with controlled unitary has been proposed in \cite{Lu2023}, where an initial $\alpha=2$ cluster SPT state is turned into a {\it non-glassy}  Ising-type GHZ state on even sites through local measurements on odd sites and feedback operation. 
Here, we shall give an extension of that feedback unitary for the generic $\alpha$ systems. 

\noindent\underline{Feedback unitary for each measurement step}: 
We discuss the extended feedback at $(m_s=k+1)$-th measurement step with the outcomes $\vec{\beta}^k$ denote by $U^f_{k}({\vec\beta}^k)$, explicit form of which is given as
\begin{eqnarray}
U^f_{k}(\vec{\beta}^k)&=&\prod^{\alpha-1}_{m=k+1}u^m(\vec{\beta}^k),\\
u^m(\vec{\beta}^k)&=&
\prod^{L/\alpha-1}_{\ell=0}X^{\frac{1-\prod^{\ell}_{q=0}\beta_{\alpha q+k}}{2}}_{\alpha \ell+m}.
\label{feedback_uni}
\end{eqnarray}
Thus, controlled unitary at $(m_s=k+1)$-th measurement step, $U^c_k$,  is defined as composite of the following operators;
\begin{eqnarray}
U^c_{k}({\vec\beta}^k)=U^f_{k}({\vec\beta}^k)P^{k}_{{\vec \beta}^k},
\label{control_U}
\end{eqnarray}
\begin{eqnarray}
    U^c_k \equiv \sum_{\vec{\beta}^k}U^c_{k}({\vec\beta}^k).
    \label{Uck}
\end{eqnarray}
The form of this controlled unitary $U^c_k$ can be regarded as an extended version proposed in \cite{Tantivasadakarn2022,Lu2023}. 

Then, following Ref.~\cite{Tantivasadakarn2022}, we consider sequential measurements with the outcome feedback, starting from a generic even $\alpha$ cluster SPT state as an example. The first step of the controlled unitary is 
\begin{eqnarray}
U^c_{0}({\vec\beta}^0)|CS_e(\alpha)\rangle \propto |CS_o(\alpha-1)\rangle \otimes |\vec{\beta}^0_x\rangle_{(sys)_{0}}.
\label{1st_md_state_feed}
\end{eqnarray}
On the unmeasured sites, we obtain {\it non-glassy} $(\alpha-1)$ cluster SPT state denoted by $|CS_o(\alpha-1)\rangle$ with $P_0=+1$ corresponding to the positive-parity ground state of the {\it original Hamiltonian} 
$H_{\rm gc}(\alpha-1)$ in Eq.~(\ref{gc_model}). 

By using this procedure for $(\alpha-1)$ times to the initial state,
the non-glassy reduced cluster SPT states on the unmeasured sites emerge at each step. 
By the same procedure explained in Sec.IV A, after ($m_s=\alpha-1$) steps by the controlled feedback $U^c_{k}(\vec{\beta}^k)$, we finally obtain the state in the following form:
\begin{eqnarray}
\biggl(\prod^{\alpha-2}_{k=0} U^c_{k}(\vec{\beta}^k)\biggr)|CS_e(\alpha)\rangle \propto |{\rm GHZ}_+\rangle_{(sys)_{\alpha-1}} \bigotimes^{\alpha-2}_{k=0} |\vec{\beta}^{k}_x\rangle_{(sys)_{k}}.\nonumber\\
\label{last_md_state_feed}
\end{eqnarray}
We obtain the $\alpha$-period LRO state as the {\it clean and no-glassy} GHZ state with $P_{\alpha-2}=+1$ denoted by $|{\rm GHZ}^g_+\rangle_{(sys)_{\alpha-1}}$ on the subsystem $(sys)_{\alpha-1}$. 
After all, $(\alpha-1)$-times controlled-unitary operations are applied to the $\alpha$ cluster SPT state. 
At each step, we obtain the reduced non-glassy cluster SPT state on the unmeasured subsystem and finally get the clean GHZ state, having $\alpha$-period LRO.
A schematic image of this procedure is shown in Fig.~\ref{Fig2}(b).
Obviously, this manipulation is also applicable for the general odd $\alpha$-cases.\\

\noindent\underline{Mixed-state picture}:
The above procedure is discussed in the purified picture as in Ref.~\cite{Tantivasadakarn2022}. 
As proposed in Ref.~\cite{Lu2023}, the manipulation under consideration can be applied to mixed states with the local operation and classical communication (LOCC). 

We rewrite the initial state $|\Psi^{\alpha}_0\rangle=|CS(\alpha)\rangle$ in terms of its density matrix $\rho^{\alpha}_0=|\Psi^{\alpha}_0\rangle \langle \Psi^{\alpha}_0|$, 
where we consider one of the two-fold degenerate ground states of $H_{\rm gc}(\alpha)$ with $P=+1$ for odd $\alpha$ case,
and apply the first step ($m_s=1$) of the controlled unitary $U^c_{0}(\vec{\beta}^0)$ to $\rho^{\alpha}_0$~\cite{Lu2023}, then
the density matrix changes as 
\begin{eqnarray}
\rho^{\alpha}_1=\sum_{\vec{\beta}^{0}}U^c_{0}(\vec{\beta}^0)\rho^{\alpha}_0U^{c\dagger}_{0}(\vec{\beta}^0).
\end{eqnarray}
The mixed state after the measurement exhibits the order of the $(\alpha-1)$ cluster SPT state. 

If this approach is repeated by $m_s$ times, we obtain a mixed state after $m_s$ measurement steps with the feedback, denoted by $\rho^{\alpha}_{m_s}$. 
The mixed state $\rho^{\alpha}_{m_s}$ has the string order of the $(\alpha-{m_s})$ cluster SPT state. 
We can analytically prove this observation from the finite string order of $\alpha$ cluster SPT state as follows,
\begin{eqnarray}
&&1=\langle CS(\alpha)|\hat{S}(\alpha,\alpha i_0+m_s,\alpha k_0+m_s)|CS(\alpha)\rangle\nonumber\\
&&=\mathrm{tr}\biggr[\hat{S}(\alpha-m_s,\alpha i_0+m_s,\alpha k_0+m_s)\rho^\alpha_{m_s}\biggl].
\label{final_claim}
\end{eqnarray}
The explicit form of the $\alpha'$-string order operator $\hat{S}(\alpha',\alpha i_0+m,\alpha k_0+m)$ is given by Eq.~(\ref{STO}) below.
The proof of the above equation is given in Appendix. D and E. 

Finally, we repeat this procedure $m_s=(\alpha-1)$ times and obtain the following mixed state density matrix,
\begin{eqnarray}
\rho^{\alpha}_{\alpha-1}=\sum_{\vec{\beta}^{0},\cdots,\vec{\beta}^{\alpha-2} }\biggl(U^c_{\alpha-2}(\vec{\beta}^{\alpha-2}) \cdots U^c_{0}(\vec{\beta}^0)\biggr)\rho^{\alpha}_0\nonumber\\
\times\biggl(U^{c\dagger}_{0}(\vec{\beta}^{0}) \cdots U^{c\dagger}_{\alpha-2}(\vec{\beta}^{\alpha-2})\biggr).
\end{eqnarray}
We expect that this density matrix, $\rho^{\alpha}_{\alpha-1}$ 
exhibits the following LRO of the Ising GHZ-type such as 
\begin{eqnarray}
\mathrm{tr}[\rho^{\alpha}_{\alpha-1}Z_{i_1} Z_{i_2}]=1,
\end{eqnarray}
where $i_1$ and $i_2$ are sites of the subsystem $(sys)_{\alpha-1}$.

\section{Numerical verification without feedback unitary by using the efficient stabilizer simulation}

We have given the qualitative discussion and concrete examples of the measurement-reduction hierarchy starting from generalized $\alpha$ cluster SPT states. 
In what follows, we numerically show evidences of the emergent hierarchy structure 
(for the systems without the feedback unitary), the numerical calculation of which can be performed by using the efficient numerical algorithm for the stabilizer formalism \cite{Gottesman1997,Aaronson2004}.

In the numerics, we observe the following quantities. 
The first one is an extended glassy string order.
We expect that the glassy $\alpha'$ cluster SPT state can be captured by the following operator \cite{Morral-Yepes2023}
\begin{eqnarray}
\hat{S}(\alpha',i_0,k_0)=Z_{\alpha' i_0}\biggl[\prod^{k_0-1}_{i=i_0}
\biggl(\prod^{\alpha'-1}_{m=1}X_{\alpha' i+m} \biggr)\biggr]Z_{\alpha' k_0}.
\label{STO}
\end{eqnarray}
Here, please note that the supports of all the operators reside on the unmeasured sites. 
The labels are defined by $j^{m_s}$ after $m_s$ measurement steps.
As $\hat{S}(\alpha',i_0,k_0)$ takes positive and negative values randomly reflecting random outcomes,
we calculate the squared expectation value of $\hat{S}(\alpha',i_0,k_0)$ called the glassy string order (STO),
\begin{eqnarray}
S_g(\alpha',i_0,k_0)=|\langle \Psi_s|\hat{S}(\alpha',i_0,k_0)|\Psi_s\rangle|^2,
\label{STO_g}
\end{eqnarray}
where $|\Psi_s\rangle$ denotes the SPT states appearing after measurements. 
This quantity is obtained by checking the commutativity for the stabilizer generators of the state $|\Psi_s\rangle$, and does not depend on the pattern of the outcomes of measurements, the technical aspect is explained in \cite{KI2023}. 
When the state $|\Psi_s\rangle$ is in a (fixed point) glassy $\alpha'$ cluster SPT phase, $S_g(\alpha',i_0,k_0)=1$ for any $i_0$ and $k_0$.
On the other hand for state $|\Psi_s\rangle$ not in that phase, $S_g(\alpha',i_0,k_0)=0$. 

As the second quantity, we consider the following connected spin-glass long-range order parameter (SGO) \cite{Sang2021,KOI2024} 
\begin{eqnarray}
{\rm SG}(i_0,k_0)&=&|\langle \Psi_s|Z_{i_0}Z_{ k_0}|\Psi_s\rangle|^2\nonumber\\
&&-|\langle \Psi_s|Z_{i_0}|\Psi_s\rangle|^2\
-|\langle \Psi_s|Z_{k_0}|\Psi_s\rangle|^2.
\label{SGO}
\end{eqnarray}
Here, note that $i_0$, and $k_0$ are both the unmeasured sites, i.e.,
we are interested in the correlations in the subsystem $(sys)_{\alpha-1}$.
The SGO characterizes the glassy GHZ phase similar to the spin-glass-ordered phase. The numerical technical aspect is explained in \cite{Sang2021}.
When the state $|\Psi_s\rangle$ is a perfect-glassy GHZ state, then ${\rm SG}(i_0,k_0)=1$ for any $i_0$ and $k_0$. 
On the other hand for state $|\Psi_s\rangle$ not in the GHZ phase, ${\rm SG}(i_0,k_0)=0$. 

\begin{figure}[t]
\begin{center} 
\vspace{0.5cm}
\includegraphics[width=8.8cm]{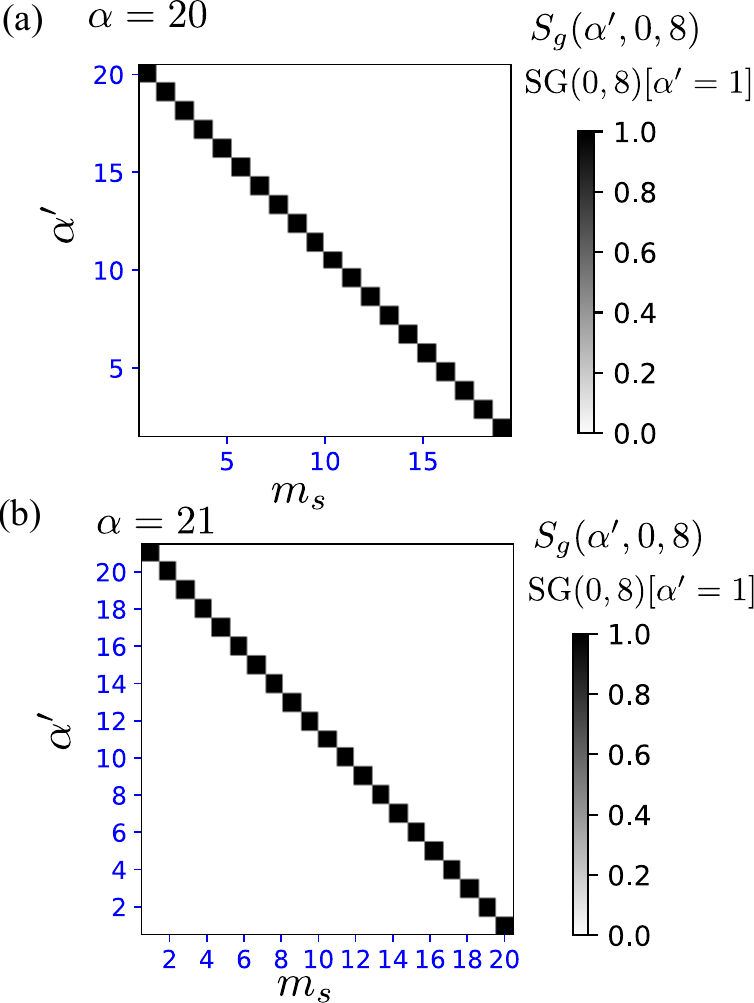}  
\end{center} 
\caption{The values of $\alpha$-glassy STO and SGO ($\alpha '=1$), (a) $\alpha=20$, $N=20$ (b) $\alpha=21$, $N=20$.
All order parameters are calculated by employing the relabeled sites defined on the unmeasured sites at each measurement step $m_s$.}
\label{Fig_order}
\end{figure}
The third quantity is the entanglement entropy for a subsystem $X$ denoted by $S_X$.
In the stabilizer formalism, the entanglement entropy is related to the number of linearly-independent stabilizers in a target subsystem $X$ \cite{Fattal2004,Nahum2017}. 
It is given by $S_X=g_X-L_X$,
where $L_X$ is the system size of the subsystem $X$,
$g_X$ is given by $\mathrm{rank}|M_X|$, where 
the matrix $M_X$ is obtained by stacking binary-represented vectors of $L$ stabilizer generators, which are spatially truncated within subsystem $X$. 
In this work, we set $X$ to a connected half subsystem including $L/2$ sites of the entire system, $L_X=L/2$. 

We turn to the numerical setup and results. 
We prepare $\alpha=20$ and $21$ cluster SPT states as an initial stabilizer state. 
We apply total $\alpha-1$ measurement steps, where we apply the measurement $
P^{m_s-1}_{\vec \beta^{m_s-1}}$ at each measurement step labeled by $m_s$, ($m_s=1, \cdots, \alpha-1$, where $m_s=0$ corresponds to no measurement to the system, that is, the system is in the initial state.) 
Please note that in the calculation of $\alpha '$-STO and SGO at each $m_s$, we consider only the unmeasured sites and calculate the $\alpha '$-STO and SGO defined on the unmeasured sites labeled by $j^{m_s}$ as shown in Fig.~\ref{Figlabel} (b) and Appendix C, that is, no measured sites are included in the definition of the operators. 
Under this prescription of the site choice, we set $i_0=0$ and $k_0=8$.

Figure \ref{Fig_order} displays results of various $\alpha '$-STO and SGO [which is nothing but the $\alpha '=1$ case] along the measurement step $m_s$. 
For even $\alpha=20$ case [Fig.~\ref{Fig_order}(a)], we see that at $m_s=0$, $S_g(\alpha'=20,i_0,k_0)=1$ and the others have zero values. 
Then, as increasing $m_s$, we observe that $S_g(\alpha'=\alpha-m_s,i_0,k_0)=1$ and the others have zero value. 
This indicates that at the measurement step $m_s$, a glassy $\alpha-m_s$ cluster SPT state emerges in the unmeasured subsystems. 
At the final step $m_s=\alpha-1$, we observe the emergence of a strict glassy GHZ state in the subsystem $(sys)_{\alpha-1}$ due to ${\rm SG}(i_0,k_0)=1$. 

For the odd $\alpha=20$ case [Fig.~\ref{Fig_order}(b)], we observe the same behavior as that of the even $\alpha$ case. 
Starting from $\alpha$ cluster SPT state, the state reaches the final glassy GHZ state through $(\alpha-1)$ sequential measurements by $
P^{m_s-1}_{\vec \beta^{m_s-1}}$.

These numerical results corroborate the argument in the previous section. 

\begin{figure}[t]
\begin{center} 
\vspace{0.5cm}
\includegraphics[width=8cm]{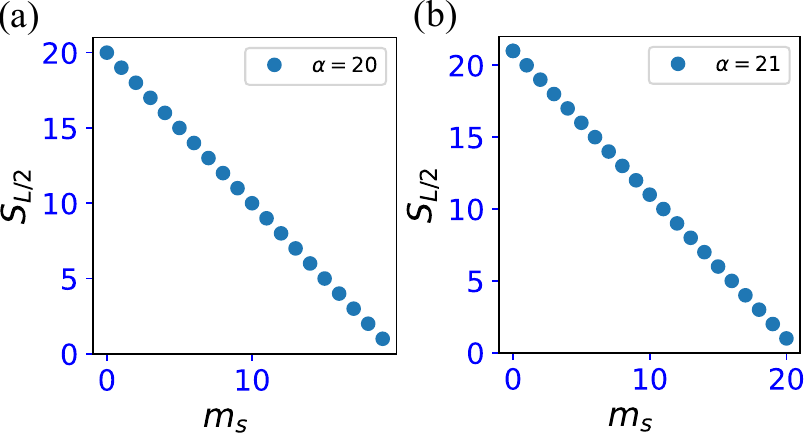}  
\end{center} 
\caption{Half subsystem entanglement entropy $S_{L/2}$: (a) $\alpha=20$, $N=20$ and the total number of the measurement steps is $19$.  (b) $\alpha=21$, $N=20$. 
The total number of the measurement steps is $20$.}
\label{Fig_EE}
\end{figure}

We finally numerically study the entanglement entropy (EE) $S_{L/2}$.
The results are shown in Fig.~\ref{Fig_EE}. 
For even $\alpha=20$ case [Fig.~\ref{Fig_EE}(a)], 
at $m_s=0$, the initial EE is $S_{L/2}=20$, which agrees with the properties of the  even $\alpha$ cluster SPT state \cite{Verresen2017}. 
We further observe the linear decreasing behavior of the EE indicating that $(\alpha-m_s)$ cluster SPT state is produced by the measurements. 
Finally at $m_s=\alpha-1$, we see $S_{L/2}=1$, indicating the emergence of the glassy GHZ state \cite{Lang2020,KOI2024}.

For the odd $\alpha=21$ case [Fig.~\ref{Fig_EE}(b)], 
at $m_s=0$, the initial EE is $S_{L/2}=21$, as expected for the odd $\alpha$ cluster SPT state \cite{Verresen2017}. 
We further observe the linear decreasing behavior of the EE similar to that of the even $\alpha$ case. 
Finally at $m_s=\alpha-1$, we also see $S_{L/2}=1$. 

Here, we comment that the value of $S_{L/2}$ is related to the number of emerging edge modes when we introduce a cut of the system or employ open boundary conditions \cite{Verresen2017}. In addition, the study of the topological response such as the topological pump and edge theory for the generalized cluster model is interesting. Recently, such a direction of study appeared \cite{Fechisin2023}.
A study concerning edge modes in the present setup is a future problem. 

\section{Conclusion}
In this study, we have proposed a measurement-reduction hierarchy of the generalized cluster SPT state by sequential subsystem projective measurements from an initial generalized cluster SPT state with large symmetry. 
We expect that only the sequential subsystem projective measurements induce the series of the glassy cluster SPT states and finally a glassy GHZ state. Furthermore, we found efficient feedback unitary regarded as an extension form to that of the previous studies \cite{Tantivasadakarn2022,Lu2023}. 
The mixed states created by the sequential controlled unitary exhibit an extended string order at each measurement step indicating the emergence of the reduced cluster SPT state on the unmeasured sites.
The class of the cluster SPT state depends on the number of the measurement step.

We expect that our investigated scheme of sequential measurements for a particular symmetry generator has broad applications to various SPTs defined on high-dimensional systems with a number of large symmetry groups.
There, the reduction procedure can act properly. 
It has already been shown that at one-layer level of subsystem measurements a 2D cluster SPT turns into a 2D LRO state \cite{Tantivasadakarn2022} and an long-range entangled state can be produced \cite{Lu2022_PRX3}. 

Further, how the bulk measurement affects the topological response and edge theory in the generalized cluster model \cite{Fechisin2023} is also interesting. Such a study have potential to give a new avenue in this research fields. 

To apply the findings in this work to practical systems of quantum information etc, more detailed study on global and topological properties of the extended cluster models is welcome.
One interesting and promising direction is a gauging method of the model, which produces a novel and broad class of duality.
This prescription has already been applied to the $ZXZ$ model \cite{Seifnashri2024}.
For the extended cluster models, a couple of similar methods can be considered, which render the global
$Z_2$ symmetries to local ones by introducing `gauge fields'.
Using Gauss's laws, the model Hamiltonian is expressed in terms of gauge-invariant quantities.
This is an interesting future work, and we hope that we will report fruitful results on it in the near future.

\section*{Acknowledgements}
This work is supported by JSPS KAKENHI: JP23K13026(Y.K.) and JP23KJ0360(T.O.). 

\appendix

\section*{Appendix A: Basic transformation in a set of stabilizer generators}
We consider a set of the stabilizer generators denoted by $[g_0,\cdots, g_{N-1}]$, where $N$-generators are included and they are linearly independent with each other.
As explained in Ref.~\cite{Nielsen_Chuang}, there is a standard transformation between stabilizers. 
We can freely change the set of stabilizer generators by multiplying $g_i$ by $g_j$ ($i\neq j$) to obtain a new stabilizer generator $g_{i}\to g_{i}g_j\equiv g'_{i}$. 
Under this transformation, the stabilizer group obtained from stabilizer generators is invariant. 
By using this rule including the sign of the stabilizer generators, we can construct a tractable set of stabilizer generators to identify the corresponding many-body states. 
This prescription woks similarly for the stabilizer generators with the outcome factors $g_i\to \beta_jg_{i}$ with $\beta_j=\pm 1$. 
In the standard transformation, we can change the form of the stabilizer generators by multiplying $\beta_ig_i$ with $\beta_j g_j$ ($i\neq j$) to obtain a transformed stabilizer generator as  $\beta_i g_{i}\to \beta_i\beta_jg_{i}g_j\equiv \beta_i\beta_j g'_{i}$.

\section*{Appendix B: Update rule of projective measurements in efficient numerical algorithm of stabilizer formalism}
We review a simple update procedure for a projective measurement in Aaronson-Gottesman efficiently stabilizer algorithm \cite{Gottesman1997,Aaronson2004,Nielsen_Chuang}. 
This update method is efficient not only for numerical calculations but also for the analytical calculation to deduce a measured many-body state generated by projective measurements with Pauli string measurement operators.

We employ a sightly different notation to include sign of outcomes \cite{Gottesman1997,Aaronson2004,Nielsen_Chuang}. 
This notation is useful to write down an effective Hamiltonian after measurements and to elucidate the relations among values of outcomes in distinct sites.

The efficient update way is as follows:\\
Suppose that a pure state in a $N$-qubit system is stabilized by a set of $N$ stabilizer generators. 
We denote this set by $\mathcal{S}=[g_0,g_1,\cdots,g_{N-1}]$, and call the state under consideration stabilizer state ${\cal S}$.
For this stabilizer state $\mathcal{S}$, let us measure the physical quantity corresponding to the operator $s$ in a Pauli group $\mathcal{P}_{N}$ (Pauli string operators), where we consider $s^2=1$ and the outcome takes $\beta_{s}=\pm 1$. 
[In numerical calculations, we will ignore the sign of the measured value since it does not affect the result for our target physical quantities.]

Then, by the projective measurement on the state $\mathcal{S}$, the stabilizer generators are updated as follows~\cite{Gottesman1997,Aaronson2004}:
\begin{itemize}
\item[(I)] Search anticommutative stabilizer generators to $s$. 
This can be carried out by using the check matrix \cite{Nielsen_Chuang}. 
From this procedure, as the case 1, we obtain single or some $m$ anticommutative stabilizer generators, $g_{\ell_1},g_{\ell_2},\cdots,g_{\ell_m}$ ($m\leq N$). As the case 2, there is no anticommutative one, $\mathcal{S}$ is not updated.

\item[(II)] If the case 1 occurs in (I) and there is a only single stabilizer generator denoted by $g_{m_1}$ anticommute to $s$, we replace $g_{m_1}$ with $\beta_s s$. Here $\beta_s$ is the outcome with probability $p_{\beta_s}=\sqrt{\langle \Psi_{st}|P_{\beta_s}|\Psi_{st}\rangle}=1/2$ with $\displaystyle{P_{\beta_s}=\frac{1}{2}[I+\beta_s s]}$\cite{Nielsen_Chuang}, where $|\Psi_{st}\rangle$ is the stabilizer state by $\mathcal{S}$. The update of  $\mathcal{S}$ is achieved. 

\item[(III)] When there are (more than two) $m$ anticommutative stabilizer generators $g_{\ell_1},g_{\ell_2},\cdots, g_{\ell_m}$ ($m\leq N$), replace $g_{\ell_1}$ with $\beta_s s$. Furthermore, for the rest of anticommutative ones $g_{\ell_i}$, update $g_{\ell_i} \to g_{\ell_i} g_{\ell_1}$. By this procedure the set of stabilizer generators $\mathcal{S}$ is updated by the projective measurements with the measurement operator $s$.
\end{itemize}

Furthermore, when we next carry out the projective measurement with a measurement operator $s'$ with the outcome $\beta_{s'}$, we do the above update prescription again but by treating the stabilizer generators with the previous outcome factor $\beta_s$, such as $\beta_s s$.


\section*{Appendix C: Relabel functions}
In this appendix, we explain the re-numbering rule of the unmeasured sites after $m_s$ measurement steps.

First, $j$ is the initial site label as shown in Fig.~\ref{Figlabel} (b). 
Then, the site-relabeling after the first ($m_s=1$) measurement step,
\begin{eqnarray}
j^1[j]\equiv(\lfloor \frac{j}{\alpha} \rfloor)(\alpha-1)+ [(j\bmod \alpha)-1], \nonumber
\end{eqnarray}
for $j \in (all)-(sys)_0$. 
The site label $j^1$ labels correctly the unmeasured sites in order as Fig.~\ref{Figlabel} (b). 

Generally, the site-relabeling after $m_s$ measurement steps denoted as $j^{m_s}$ is given by
\begin{eqnarray}
j^{m_s}[j]\equiv(\lfloor \frac{j}{\alpha} \rfloor)(\alpha-m_s)+ [(j\bmod \alpha)-m_s], \nonumber
\end{eqnarray}
for $j \in (all)-\sum^{m_s-1}_{k=0}(sys)_k$.

Also, in the effective Hamiltonian after the first step measurement, the site-label of the outcome factor $\beta$ is given by  
\begin{eqnarray}
n^{0}[j^1]\equiv (\lfloor \frac{j^1}{\alpha-1} \rfloor+1)\alpha+0. \nonumber
\end{eqnarray}

Generally, for the effective Hamiltonian after $m_s$ measurement steps, the site-label of the outcome factor $\beta$ in the effective Hamiltonian is given by
\begin{eqnarray}
n^{m_s-1}[j^{m_s}]\equiv (\lfloor \frac{j^{m_s}}{\alpha-{m_s}} \rfloor+1)\alpha+({m_s}-1).\nonumber
\end{eqnarray}

Note that the inverse re-labeling function can be also defined for all site-labeling rules appearing here.

\widetext
\section*{Appendix D: presence of string order at any measurement step}
By extending the observation in \cite{Lu2023}, we shall prove that a series of the mixed state $\rho^{\alpha}_{m_s}$ have their own finite string order. 

We start $\alpha$ cluster SPT state. 
In the following calculation, we use the initial site label even after any measurements.

The string order for the initial state is
\begin{eqnarray}
\langle CS(\alpha)|\hat{S}(\alpha,\alpha i_0+m_s,\alpha k_0+m_s)|CS(\alpha)\rangle
=
\langle CS(\alpha)|
Z_{\alpha i_0+m_s}\biggl[\prod^{k_0-1}_{i=i_0}
\biggl(X_{\alpha i+1+m_s}\cdots X_{\alpha i+\alpha-1+m_s} \biggr)\biggr]Z_{\alpha k_0+m_s}|CS(\alpha)\rangle
=1.\nonumber\\
\label{string_alpha}
\end{eqnarray}
Here, we suitably set the sites of the string operators such that the edges of sites are set as $\alpha i_0+m_s$ and $\alpha k_0+m_s$. 
$m_s$ is the target number of the measurement steps, $m_s$, where we consider $1\leq m_s\leq \alpha-1$.

The above string order becomes
\begin{eqnarray}
&&\langle CS(\alpha)|
Z_{\alpha i_0+m_s}\biggl[\prod^{k_0-1}_{i=i_0}
\biggl(X_{\alpha i+1+m_s}\cdots X_{\alpha i+\alpha-1+m_s} \biggr)\biggr]Z_{\alpha k_0+m_s}|CS(\alpha)\rangle\nonumber\\
&&=\sum_{\vec{\beta}^0,\cdots,\vec{\beta}^{m_s-1}}
\langle CS(\alpha)|\biggl[P^{0}_{\vec{\beta}^0}\cdots P^{m_s-1}_{\vec{\beta}^{m_s-1}}\biggr]
Z_{\alpha i_0+m_s}\biggl[\prod^{k_0-1}_{i=i_0}
\biggl(X_{\alpha i+1+m_s}\cdots X_{\alpha i+\alpha-1+m_s} \biggr)
\biggr]Z_{\alpha k_0+m_s}\biggl[P^{m_s-1}_{\vec{\beta}^{m_s-1}}\cdots P^{0}_{\vec{\beta}^{0}}\biggr]|CS(\alpha)\rangle\nonumber\\
&&=\sum_{\vec{\beta}^0,\cdots,\vec{\beta}^{m_s-1}}
\langle CS(\alpha)|\biggl[P^{0}_{\vec{\beta}^0}\cdots P^{m_s-1}_{\vec{\beta}^{m_s-1}}\biggr]
Z_{\alpha i_0+m_s}\biggl[\prod^{k_0-1}_{i=i_0}
\biggl(X_{\alpha i+1+m_s}\cdots X_{\alpha i+\alpha-1} \biggr)
\biggr]
\biggl[\prod^{k_0}_{i=i_0+1}
\biggl(X_{\alpha i}\cdots X_{\alpha i+m_s-1} \biggr)
\biggr]\nonumber\\
&&\times 
Z_{\alpha k_0+m_s}\biggl[P^{m_s-1}_{\vec{\beta}^{m_s-1}}\cdots P^{0}_{\vec{\beta}^{0}}\biggr]|CS(\alpha)\rangle\nonumber\\
&&=\sum_{\vec{\beta}^0,\cdots,\vec{\beta}^{m_s-1}}
\langle CS(\alpha)|\biggl[P^{0}_{\vec{\beta}^0}\cdots P^{m-1}_{\vec{\beta}^{m_s-1}}\biggr]
Z_{\alpha i_0+m_s}\biggl[\prod^{k_0-1}_{i=i_0}
\biggl(X_{\alpha i+1+m_s}\cdots X_{\alpha i+\alpha-1} \biggr)
\biggr]
\biggl[\prod^{k_0}_{i=i_0+1}
\biggl(\beta_{\alpha i}\cdots \beta_{\alpha i+m_s-1} \biggr)
\biggr]\nonumber\\
&&\times 
Z_{\alpha k_0+m_s}\biggl[P^{m_s-1}_{\vec{\beta}^{m_s-1}}\cdots P^{0}_{\vec{\beta}^{0}}\biggr]|CS(\alpha)\rangle\nonumber\\
&&=\sum_{\vec{\beta}^0,\cdots,\vec{\beta}^{m_s-1}}
\langle CS(\alpha)|\biggl[P^{0}_{\vec{\beta}^0}\cdots (P^{m_s-1}_{\vec{\beta}^{m_s-1}}U^{f\dagger}_{m_s-1}({\vec\beta}^{m_s-1}))\biggr]
Z_{\alpha i_0+m_s}\biggl[\prod^{k_0-1}_{i=i_0}
\biggl(X_{\alpha i+1+m_s}\cdots X_{\alpha i+\alpha-1} \biggr)
\biggr]
\biggl[\prod^{k_0}_{i=i_0+1}
\biggl(\beta_{\alpha i}\cdots \beta_{\alpha i+m_s-2} \biggr)
\biggr]\nonumber\\
&&\times 
Z_{\alpha k_0+m_s}\biggl[(U^{f}_{m_s-1}({\vec\beta}^{m_s-1})P^{m_s-1}_{\vec{\beta}^{m_s-1}})\cdots P^{0}_{\vec{\beta}^{0}}\biggr]|CS(\alpha)\rangle,
\label{string_alpha2}
\end{eqnarray}
where in the second line, we have used $\sum_{\vec{\beta}^k} P^{k}_{\vec{\beta}^{k}}=\sum_{\vec{\beta}^k} (P^{k}_{\vec{\beta}^{k}})^2=1$ and in the last line
\begin{eqnarray}
U^{f\dagger}_{m_s-1}({\vec\beta}^{m_s-1})Z_{\alpha i_0+m_s}
(X\cdots X)Z_{\alpha k_0+m_s}U^{f\dagger}_{m_s-1}({\vec\beta}^{m_s-1})
=Z_{\alpha i_0+m_s}
(X\cdots X)\biggl[\prod^{k_0}_{i=i_0+1}\beta_{\alpha i+m_s-1}\biggr]Z_{\alpha k_0+m_s}.
\label{important_formula}
\end{eqnarray}
The proof of this equation is given in Appendix.E. 

We further proceed the calculation from Eq.~(\ref{string_alpha2}),
\begin{eqnarray}
{\mbox{Eq.~(\ref{string_alpha2})}}
&=&\sum_{\vec{\beta}^0,\cdots,\vec{\beta}^{m_s-1}}
\langle CS(\alpha)|\biggl[(P^{0}_{\vec{\beta}^0}U^{f\dagger}_{0}({\vec\beta}^{0}))
\cdots (P^{m_s-1}_{\vec{\beta}^{m_s-1}}U^{f\dagger}_{m_s-1}({\vec\beta}^{m_s-1}))\biggr]
Z_{\alpha i_0+m_s}\biggl[\prod^{k_0-1}_{i=i_0}
\biggl(X_{\alpha i+1+m_s}\cdots X_{\alpha i+\alpha-1} \biggr)
\biggr]\nonumber\\
&&\times 
Z_{\alpha k_0+m_s}\biggl[(U^{f}_{m_s-1}({\vec\beta}^{m_s-1})P^{m_s-1}_{\vec{\beta}^{m_s-1}})\cdots (U^{f}_{0}({\vec\beta}^{0})P^{0}_{\vec{\beta}^{0}})\biggr]|CS(\alpha)\rangle\nonumber\\
&&=\sum_{p}\sum_{\vec{\beta}^0,\cdots,\vec{\beta}^{m_s-1}}
\langle CS(\alpha)|\biggl[(P^{0}_{\vec{\beta}^0}U^{f\dagger}_{0}({\vec\beta}^{0}))
\cdots (P^{m_s-1}_{\vec{\beta}^{m_s-1}}U^{f\dagger}_{m_s-1}({\vec\beta}^{m_s-1}))\biggr]
|\psi_p\rangle\langle \psi_p|
\hat{S}(\alpha-m_s,\alpha i_0+m_s,\alpha k_0+m_s)\nonumber\\
&&\times
\biggl[(U^{f}_{m_s-1}({\vec\beta}^{m_s-1})P^{m_s-1}_{\vec{\beta}^{m_s-1}})\cdots (U^{f}_{0}({\vec\beta}^{0})P^{0}_{\vec{\beta}^{0}})\biggr]|CS(\alpha)\rangle\nonumber\\
&&=\mathrm{tr}\biggr[\hat{S}(\alpha-m_s,\alpha i_0+m_s,\alpha k_0+m_s)\rho^\alpha_{m_s}\biggl],
\label{important_relation2}
\end{eqnarray}
where
$\hat{S}(\alpha-m_s,\alpha i_0+m_s,\alpha k_0+m_s)$ is the operator of $(\alpha-m_s)$ string order and the sites on the operators are in the unmeasured sites, and $\rho^{\alpha}_{m_s}$ is 
\begin{eqnarray}
&&\rho^{\alpha}_{m_s}=\sum_{\vec{\beta}^0,\cdots,\vec{\beta}^{m_s-1}}
\biggl[U^{c}_{m_s-1}({\vec\beta}^{m_s-1})\cdots U^{c}_{0}({\vec\beta}^{0})\biggr]|CS(\alpha)\rangle
\langle CS(\alpha)|\biggl[U^{c\dagger}_{0}({\vec\beta}^{0})
\cdots U^{c\dagger}_{m_s-1}({\vec\beta}^{m_s-1})\biggr].
\label{density_matrix_m-1}
\end{eqnarray}
We have used the completeness relation for a set of $L$-site qubit orthogonal basis, $\sum_{p}|\psi_p\rangle\langle \psi_p|=1$.

From this calculation, from the presence of the string order of the initial $\alpha$ cluster SPT state, we conclude that the measured and feedbacked state after $m_s$ times one-layer measurements for each different subsystem $(sys)_{k}$ for $k=0,\cdots ,m_s-1$ also has $(\alpha-m_s)$ string order, 
\begin{eqnarray}
&&1=\langle CS(\alpha)|\hat{S}(\alpha,\alpha i_0+m_s,\alpha k_0+m_s)|CS(\alpha)\rangle=\mathrm{tr}\biggr[\hat{S}(\alpha-m_s,\alpha i_0+m_s,\alpha k_0+m_s)\rho^\alpha_{m_s}\biggl].
\label{final_claim}
\end{eqnarray}
From this relation, we expect the presence of the string order for any measurement step except for $(\alpha-1)$ step. 
This indicates that the cluster SPT state on unmeasured sites exists for any measurement step except for $(\alpha-1)$ step and the class of the string order depends on the numbers of the measurement step $m_s$. 
This relation nothing but indicates a measurement reduction hierarchy. 
Also, Eq.(42) for $m_s=\alpha-1$ case is also satisfied, corresponding to the Ising GHZ LRO.

\section*{E. Proof of Eq.~(\ref{important_formula})}
We here show the proof of Eq.~(\ref{important_formula}).
\begin{eqnarray}
&&U^{f}_{m_s-1}({\vec\beta}^{m_s-1})Z_{\alpha i_0+m_s}
(X\cdots X)Z_{\alpha k_0+m_s}U^{f\dagger}_{m_s-1}({\vec\beta}^{m_s-1})
\nonumber\\
&&=
U^{f}_{m_s-1}({\vec\beta}^{m_s-1})Z_{\alpha i_0+m_s}U^{f\dagger}_{m_s-1}({\vec\beta}^{m_s-1})
(X\cdots X)U^{f}_{m_s-1}({\vec\beta}^{m_s-1})Z_{\alpha k_0+m_s}U^{f\dagger}_{m_s-1}({\vec\beta}^{m_s-1}).
\label{important_formula_p1}
\end{eqnarray}
Here, 
\begin{eqnarray}
&&U^{f}_{m_s-1}({\vec\beta}^{m_s-1})Z_{\alpha i_0+m_s}U^{f\dagger}_{m_s-1}({\vec\beta}^{m_s-1})=u^{m_s}(\vec{\beta}^{m_s-1})Z_{\alpha i_0+m_s}u^{m_s\dagger}(\vec{\beta}^{m_s-1})\nonumber\\
&&=\biggl(X_{\alpha i_0+m_s}^{\frac{1-\prod^{i_0}_{q=0}\beta_{\alpha q+(m_s-1)}}{2}}\biggr)
Z_{\alpha i_0+m_s}
\biggl(X_{\alpha i_0+m_s}^{\frac{1-\prod^{i_0}_{q=0}\beta_{\alpha q+(m_s-1)}}{2}}\biggr)
=\biggl[\prod^{i_0}_{q=0}\beta_{\alpha q+(m_s-1)}\biggr]Z_{\alpha i_0+m_s},
\nonumber\\
&&U^{f}_{m_s-1}({\vec\beta}^{m_s-1})Z_{\alpha k_0+m_s}U^{f\dagger}_{m_s-1}({\vec\beta}^{m_s-1})=\biggl[\prod^{k_0}_{q=0}\beta_{\alpha q+(m_s-1)}\biggr]Z_{\alpha k_0+m_s}.
\label{important_formula_p2}
\end{eqnarray}
Thus, by substituting the above equations into Eq.~(\ref{important_formula_p1}), we obtain
\begin{eqnarray}
\mbox{Eq.~(\ref{important_formula_p1})}=\biggl[\prod^{i_0}_{q=0}\beta_{\alpha q+(m_s-1)}\biggr]Z_{\alpha i_0+m_s}
(X\cdots X)\biggl[\prod^{k_0}_{q=0}\beta_{\alpha q+(m_s-1)}\biggr]Z_{\alpha k_0+m_s}
=Z_{\alpha i_0+m_s}
(X\cdots X)\biggl[\prod^{k_0}_{i=i_0+1}\beta_{\alpha i+m_s-1}\biggr]Z_{\alpha k_0+m_s}.\nonumber\\
\label{important_formula_proof}
\end{eqnarray}

\endwidetext

\end{document}